\documentclass[english,prl,twocolumn]{revtex4}
\usepackage{mathpazo}
\usepackage[T1]{fontenc}
\usepackage[latin9]{inputenc}
\usepackage{textcomp}
\usepackage{amsthm}
\usepackage{amsmath}
\usepackage{amssymb}

\makeatletter
\@ifundefined{textcolor}{}
{%
 \definecolor{BLACK}{gray}{0}
 \definecolor{WHITE}{gray}{1}
 \definecolor{RED}{rgb}{1,0,0}
 \definecolor{GREEN}{rgb}{0,1,0}
 \definecolor{BLUE}{rgb}{0,0,1}
 \definecolor{CYAN}{cmyk}{1,0,0,0}
 \definecolor{MAGENTA}{cmyk}{0,1,0,0}
 \definecolor{YELLOW}{cmyk}{0,0,1,0}
}

\usepackage{mathpazo}
\usepackage{amsfonts}

\usepackage{latexsym}

\makeatother
\newtheorem{thm}{Theorem}\newtheorem{cor}{Corollary}\newtheorem{lem}{Lemma} \newtheorem{exm}{Example}

\newtheorem{defn}{Definition}

\def\ket{\rangle}\def\be{\begin{eqnarray}}\def\ee{\end{eqnarray}}\def\bee{\begin{eqnarray*}}\def\eee{\end{eqnarray*}}\def\ot{\otimes}\def\tr{\hbox{Tr}}\def\qed{{\bf QED}}

\makeatother

\makeatother

\makeatother

\makeatother

\makeatother

\makeatother

\usepackage{babel}

\makeatother

\usepackage{babel}

\makeatother

\usepackage{babel}

\makeatother

\usepackage{babel}

\makeatother

\usepackage{babel}

\makeatother

\usepackage{babel}

\makeatother

\usepackage{babel}

\makeatother

\usepackage{babel}

\makeatother

\usepackage{babel}
\begin{document}

\title{Entanglement as a resource for local state discrimination in multipartite systems}

\author{Somshubhro Bandyopadhyay}

\email{som@jcbose.ac.in}

\selectlanguage{english}%

\author{Saronath Halder}

\affiliation{Department of Physics and Center for Astroparticle Physics and Space
Science, Bose Institute, Block EN, Sector V, Bidhan Nagar, Kolkata
700091}

\author{Michael Nathanson}

\email{man6@stmarys-ca.edu }

\selectlanguage{english}%

\affiliation{Department of Mathematics and Computer Science, Saint Mary's College
of California, Moraga, CA, 94556, USA}
\begin{abstract}
We explore the question of using an entangled state as a universal resource for
implementing quantum measurements by local operations and classical
communication (LOCC). We show that for most systems consisting
of three or more subsystems, there is no entangled state from the
same space that can enable all measurements by LOCC. This is in direct
contrast to the bipartite case, where a maximally entangled state
is an universal resource. Our results are obtained showing an equivalence between
the problem of local state transformation and that of entanglement-assisted
local unambiguous state discrimination.  
\end{abstract}

\maketitle

\section{Introduction}
Given a quantum system composed of spatially-separated subsystems,
it is well known that the set of quantum operations which can be implemented
with local operations and classical communication (LOCC) constitutes
a strict subset of all quantum operations on the whole system. For
example, it is impossible, by LOCC, to transform a product state into
an entangled state \cite{Entanglement-horodecki}, even with nonzero
probability. The presence of additional entanglement, however, can help to overcome
such restrictions \cite{BBKW-2009,Berry-2007,BRW-2010,Cirac-et-al-2001,Cohen-2008,Collins-et-al-2001};
and appropriate shared entanglement enables local implementation of
any quantum operation.
In this way, entanglement can be seen as a resource for quantum operations,
e.g., quantum teleportation \cite{Teleportation}, superdense coding
\cite{Densecoding}, entanglement-catalysis \cite{Jonathan-Plenio-1999},
entangling measurements and unitaries \cite{BBKW-2009,Berry-2007,BRW-2010,Cirac-et-al-2001,Collins-et-al-2001}. 

In this work, we explore the question of {\it universal} resource states, whose presence overcomes the limitations imposed by LOCC to allow implementation of whole classes of quantum operations. For instance, in a bipartite quantum system $\mathcal{H}=\mathcal{H}_A \ot \mathcal{H}_B$, a maximally entangled state $\left|\Psi\right\rangle \in\mathcal{H}$ constitutes a universal resource for extracting classical information. This can be easily understood by imagining the subsystems as controlled by two parties, Alice and Bob, who share an unknown state $\vert \phi \ket \in \mathcal{H}_A \ot \mathcal{H}_B$. If they also share the maximally entangled state $\vert \Psi \ket$, they can use it to teleport Alice's qudit to Bob. Bob can then extract information about $\vert \phi \ket$ by applying any quantum measurement to his own system, thus overcoming the limitations of the spatial separation. Because Alice and Bob can use $\vert \Psi \ket$ to perform {\it any} complete measurement on $\vert \phi \ket$, we say that $\vert \Psi \ket$ is {\it universal} for the task of complete measurement. 

A natural question to ask is whether such universal resource states also exist for multipartite systems (those consisting of three or more subsystems), where entanglement has a more complex structure \cite{DVC}. The present work considers the
task of quantum state discrimination by LOCC \cite{B-IQC-2015,Bandyo-2011,Bennett-I-99,Bennett-II-99 +Divin-2003,Calsamiglia,Duan-2009,Duan2007,fan-2005,Ghosh-2001,Ghosh-2002,Ghosh-2004,HSSH,Nathanson-2005,Peres-Wootters-1991,Virmani-2001,Walgate-2000,Walgate-2002,Watrous-2005,Yu-Duan-1,BGK} and seeks to characterize resource states which enable us to locally distinguish \textit{any} set of orthogonal states from a fixed multipartite system. We show that for a given multipartite system, a universal resource
state for local state discrimination must in general be from a larger-dimensional
space than the states themselves. 
Moreover, there exist orthonormal bases that cannot be
perfectly distinguished by LOCC using \emph{any} resource state from
the same size state space, making them good candidates for data hiding protocols. \cite{MatthewsWehnerWinter,Eggeling2002,DiVincenzo2002,Terhal2001}

We prove our results by demonstrating an equivalence between
local unambiguous state discrimination and local state transformation. Any LOCC protocol which implements one task can be adapted to achieve the other. Since multipartite entanglement is understood primarily in terms of local state transformation, our equivalence gives us tools to understand both unambiguous and perfect discrimination with LOCC. This equivalence (which follows similar results in \cite{HSSH} and others) is useful in its own right and nicely complements recent results on resources for local state transformation \cite{GCDResource2016}. This work extends earlier analysis of entanglement as a resource in {\it bipartite} systems, which often calculate how much additional entanglement is necessary to complete a certain task locally. (Early examples include quantum teleportation and entanglement catalysis for state discrimination \cite{teleport, JonathanPlenio, Entanglement-horodecki}.)
Given the complexity of quantifying the {\it amount} of  multipartite entanglement necessary for a given task, we must pay attention to the {\it quality} of the entanglement instead. 

The rest of the paper explores these questions in the following way. In Section \ref{section:Results}, we give a precise definition of a universal resource and full statements of our results, which are proven in Section \ref{Section:Proofs}. In Section \ref{section:Examples}, we illustrate our results with examples, while Section \ref{section:Conclusions} gives conclusions and directions for future work. 

\section{Statement of Results}\label{section:Results}
\subsection{Entanglement as a Resource for LOCC Measurement}
Local state discrimination problems assume that classical information
has been encoded in quantum states and seek to determine how much
of this information can be recovered locally. Formally, we suppose that
a set of spatially separated observers share a quantum system prepared
in one of a known set of pure states $\left\{ \vert \psi_i \ket \right\} \subset \otimes_{k = 1}^N\mathcal{H}_k$, each
occurring with some nonzero probability; see e.g.,
\cite{Yu-Duan-1,Watrous-2005,Walgate-2000,Walgate-2002,Ghosh-2001,Ghosh-2002,Ghosh-2004,Nathanson-2005,fan-2005}.
The question of interest is, can LOCC protocols distinguish such quantum
states as well as global measurements can?

Although optimal discrimination of {\it two} states is always possible to implement using only LOCC \cite{Walgate-2000,Virmani-2001},  local discrimination of larger sets in general is not \cite{Bennett-I-99, Ghosh-2001}. In particular, if ${\cal B}$
is a complete basis consisting of multipartite states, LOCC is \textit{never}
sufficient to distinguish its elements if ${\cal B}$ contains any
entangled states \cite{HSSH}. It may be noted that the existence
of locally indistinguishable states imply locally hidden information,
and has thus found applications in quantum cryptography primitives
such as data hiding and secret sharing \cite{MatthewsWehnerWinter,Eggeling2002,DiVincenzo2002,Terhal2001,Markham-Sanders-2008}.

On the other hand, sets of locally
indistinguishable quantum states may become distinguishable in the
presence of shared entanglement, e.g., \cite{B-IQC-2015,Cohen-2008}, and this is the phenomenon we wish to understand better. Imagine that Alice and Bob are initially in the same location prior to going their separate ways. They know that in the future, they will need to distinguish a set of bipartite states from $\mathcal{H} \ot \mathcal{H}$; but they do not know what these states will be. Nonetheless, if they prepare a maximally entangled state  $\vert \Psi \ket \in \mathcal{H}_A \ot\mathcal{H}_B $, then for {\it any} set of states $\left\{\vert \psi_i \ket \right\} \subset  \mathcal{H}_A \ot\mathcal{H}_B$, the states $\left\{\vert \Psi \ket \ot \vert \psi_i \ket  \right\}$ can be distinguished as well using only LOCC across the $A:B$ split as they can be using any other quantum operations. The maximally entangled state is a true resource in that it enables measurements which are not otherwise possible but is consumed in the process. And it is universal--Alice and Bob can prepare $\vert \Psi \ket$ without knowing what the $\{ \vert \psi_i \ket \}$ will be.  

In this work, we seek to explore this same idea in the context of multipartite systems. If Alice, Bob, and Charlie wish to jointly prepare an entangled resource state to help them distinguish unknown sets from $\mathcal{H} \ot \mathcal{H} \ot \mathcal{H}$, it is less clear what this resource state should be; and in fact depends on the type of state discrimination they are trying to achieve. This motivates the primary definition in this work. 
\begin{defn}
Let $\mathcal{H}=\otimes_{k=1}^{N}\mathbb{C}^{d_{k}}$ be a multipartite
system and let $\mathcal{H'}=\otimes_{k=1}^{N}\mathbb{C}^{d'_{k}}$
be a system with the same multipartite structure as $\mathcal{H}$. Consider  ${\cal H'} \ot  {\cal H} $ as an $N$-partite system with local subsystems $\left( \mathbb{C}^{d'_{k}} \ot \mathbb{C}^{d_{k}}\right)$, $k = 1, \ldots, N$. 

We say that a state $\vert \Phi \ket \in {\cal H'}$ is {\bf universal for local state discrimination} in ${\cal H}$ if, for every set ${\cal B} = \{ \vert \psi_i \ket\}$ of mutually orthogonal states in ${\cal H}$, the set of states $\left|\Phi\right\rangle \otimes{\cal B}\equiv\left\{ \left|\Phi\right\rangle \otimes\left|\psi_{i}\right\rangle \right\} \subset {\cal H'} \ot  {\cal H} $ can be perfectly distinguished with LOCC.

Likewise,  $\vert \Phi \ket$ is {\bf universal for local unambiguous discrimination} in ${\cal H}$ if, for every set ${\cal B} = \{ \vert \psi_i \ket\}$ of linearly independent states in ${\cal H}$, the set of states $\left|\Phi\right\rangle \otimes{\cal B}\equiv\left\{ \left|\Phi\right\rangle \otimes\left|\psi_{i}\right\rangle \right\} \subset {\cal H'} \ot  {\cal H} $ can be unambiguously distinguished with LOCC. \end{defn}

We insist that our states be mutually orthogonal for perfect state discrimination, as this is the necessary and sufficient condition for them to be perfectly distinguishable using full quantum operations. Likewise, we assume linear independence for unambiguous state discrimination (defined in the next section), since this is the necessary and sufficient condition when we are not limited by locality. \cite{Chefles-1998}

A maximally entangled bipartite state constitutes a universal
resource for both local state discrimination and local unambiguous discrimination. In order to explore the analogous question in multipartite systems, we need to develop necessary conditions for resource states relative to each task.

\subsection{Local state transformations}\label{section:Transformations}
Our first result gives a strong equivalence between local state transformation and the problem of
\textit{unambiguous state discrimination}. In unambiguous state discrimination,
definitive knowledge of the state is balanced against a probability
of definitive failure. That is, given a state $\vert \psi \ket \in  \{ \vert \psi_i \ket \}$, we either conclude with certainty that $\vert \psi\ket = \vert \psi_i \ket$ or else we receive a failure indication. As long as each  $\vert \psi_i \ket$ is detectable with positive probability, we say that the set of states can be unambiguously discriminated; it is not hard to show that this condition is equivalent to the linear independence of the $\vert \psi_i \ket$. \cite{Chefles-1998}

If we are restricted to local operations and classical communications, it can be useful to employ entanglement in the form of a resource state $\vert \Phi \ket$. We say that the set of states $\left\{ \left\vert \Phi\right\rangle \otimes\left\vert \psi_{i}\right\rangle ,1\le i\le D\right\} \in\mathcal{H}^{\prime}\otimes\mathcal{H}$
is unambiguously locally distinguishable if there exists an LOCC
measurement $\boldsymbol{\Pi}=\left\{ \Pi_{i}\right\} _{i=1}^{D+1}$
satisfying ${\displaystyle \sum_{i}\Pi_{i}={\cal I}_{\mathcal{H}^{\prime}\otimes\mathcal{H}}}$
such that for all $i,j\in\{1,2,\ldots D\}$, 
\begin{eqnarray}
\left\langle \Phi\otimes\psi_{j}\left|\Pi_{i}\right|\Phi\otimes\psi_{j}\right\rangle  & = & \delta_{ij}\epsilon_{i}\ 
\end{eqnarray}
where $\delta_{ij}$ is the Kronecker delta and the $\epsilon_{i}$ are positive constants. The outcome $\left(D+1\right)$
is \emph{inconclusive}. It turns out that the
task of using entanglement to enable local unambiguous discrimination is closely related to
that of transforming one state into another using only LOCC. 

\begin{thm}\label{thm: unambiguous} Let ${\cal B}=\left\{ \left\vert \psi_{i}\right\rangle \right\} $
be a complete basis of $\mathcal{H}=\otimes_{k=1}^{N}\mathbb{C}^{d_{k}}$,
not necessarily orthogonal. For each $i$, define $\left|\tilde{\psi_{i}}\right\rangle $
to be the unique state in $\mathcal{H}$ which is orthogonal to every $\left|\psi_{j}\right\rangle $
with $j\ne i$.

Let $\left|\Phi\right\rangle $ be a pure state in $\mathcal{H}^{\prime}=\otimes_{k=1}^{N}\mathbb{C}^{d_{k}'}$; and denote by $\left|\Phi^{*}\right\rangle $
the entrywise complex conjugate of $\left|\Phi\right\rangle $ taken
in the standard basis. Then the set of states $\left\vert \Phi\right\rangle \otimes\mathcal{B}\equiv\left\{ \left\vert \Phi\right\rangle \otimes\left\vert \psi_{i}\right\rangle \right\} \in\mathcal{H}^{\prime}\otimes\mathcal{H}$
can be unambiguously distinguished by LOCC if and only if there
exists an LOCC protocol which transforms $\left\vert \Phi^{*}\right\rangle $
 into each $\left\vert \tilde{\psi_{i}}\right\rangle $ with
positive probability. 

In particular, if we can use LOCC to unambiguously identify each $\vert \psi_i \ket$ with probability  $\epsilon_i >0$, then there exists an LOCC protocol which transforms $\left\vert \Phi^{*}\right\rangle $
 into each $\left\vert \tilde{\psi_{i}}\right\rangle $ with
 probability at least $\frac{\epsilon_i}{D}$.
\end{thm}

This theorem is proved in Section \ref{Section:Proofs} by directly showing how a local unambiguous
measurement can be used to build a local transformation protocol,
and vice versa. As an immediate consequence, we get a necessary condition for distinguishing orthonormal bases, where for each $i$,  $\vert \tilde{\psi}\ket = \vert \psi\ket$.  
\begin{cor}\label{cor: local-transformation} Let $\mathcal{B}=\left\{ \left\vert \psi_{i}\right\rangle \right\} \subset\mathcal{H}=\otimes_{i=1}^{N}\mathbb{C}^{d_{i}}$
be a complete orthonormal basis of $\mathcal{H}$, and let $\left\vert \Phi\right\rangle \in\mathcal{H}^{\prime}=\otimes_{i=1}^{N}\mathbb{C}^{d_{i}^{\prime}}$
be fixed resource state (with $d_{i}^{\prime}$ not necessarily equal
to $d_{i}$). 

If the set of states $\left\vert \Phi\right\rangle \otimes\mathcal{B}\equiv\left\{ \left\vert \Phi\right\rangle \otimes\left\vert \psi_{i}\right\rangle \right\} \in\mathcal{H}^{\prime}\otimes\mathcal{H}$
can be perfectly distinguished by LOCC then there exists a LOCC protocol
by means of which $\left\vert \Phi^{*}\right\rangle $ is converted
to each of the $\left\vert \psi_{i}\right\rangle $ with probability
$p=\frac{1}{D}$. \end{cor}

Corollary \ref{cor: local-transformation} follows immediately from the last piece of Theorem \ref{thm: unambiguous}, setting each $\epsilon_i = 1$ and observing that if each $p_i \ge \frac{1}{D}$, then this must be an equality in order for the sum of the $p_i$ to equal one. Note that if we let $\left|\Phi\right\rangle $ be a product state,
then Theorem \ref{thm: unambiguous} implies Theorem 3 of \cite{Duan2007}, which states
that, without additional entanglement, a basis ${\cal B}$ is unambiguously
distinguishable with LOCC iff each $\left|\tilde{\psi_{i}}\right\rangle $
is a product state. Similarly, Corollary \ref{cor: local-transformation}
implies the fundamental result of \cite{HSSH} that a complete basis
can be locally distinguished only if it contains only product states.

\subsection{Entanglement classes and universal resources}\label{section:Universal}
We return now to the question for finding {\it universal} resources for local state discrimination. The challenge in finding a universal resource is a 
consequence of the complex structure of multipartite entanglement, which is often described in terms of stochastic LOCC equivalence
classes.  Two states $\left|\psi_1\right\rangle $ and $\left|\psi_2\right\rangle $
are considered to be in the same SLOCC class $\mathcal{C}$ if it is possible to use LOCC to
transform the state $\left|\psi_1\right\rangle $ into $\left|\psi_2\right\rangle $
with a positive probability of success, and also to effect the transformation
$\left|\psi_2\right\rangle \rightarrow\left|\psi_1\right\rangle $. The
SLOCC equivalence classes form a partially ordered set, with ${\cal C}_{1}\preceq{\cal C}_{2}$
if the transformation $\left|\phi_{2}\right\rangle \rightarrow\left|\phi_{1}\right\rangle $
is possible with LOCC for $\left|\phi_{i}\right\rangle \in{\cal C}_{i}$.

For a given bipartite system $\mathcal{H} = \mathcal{H}_A \ot \mathcal{H}_B$, the SLOCC partial ordering is characterized
by the existence of least upper bounds. Even though there are pairs
of states which are locally incompatible ($\vert \phi_1 \ket \not\rightarrow \vert \phi_2 \ket$ and $\vert \phi_2 \ket \not\rightarrow \vert \phi_1 \ket$) when $d>2$,  there always exists a unique SLOCC class ${\cal C}_{{\rm max}}$ of
 maximally-entangled states such that if $\left|\phi\right\rangle \in{\cal C}_{{\rm max}}$,
then the local transformation $\left|\phi\right\rangle \rightarrow\left|\psi\right\rangle $
is possible for all $\left|\psi\right\rangle \in\mathcal{H}$ (and is, in fact, possible with probability 1). 

An SLOCC class of $\mathcal{H}$ is maximal if it is not reached from
any other SLOCC class of $\mathcal{H}$. Equivalently, we say that
${\cal C}$ is maximal if, for all $\left|\phi\right\rangle \in{\cal C}$
and $\left|\psi\right\rangle \in\mathcal{H}$, $\left|\psi\right\rangle \rightarrow\left|\phi\right\rangle $
implies that $\left|\psi\right\rangle \in{\cal C}$. We refer to states in maximal SLOCC classes as {\it maximally entangled}. Most multipartite systems have more than one maximally entangled equivalence
class \cite{DVC, GW-2013}; combined with Theorems \ref{thm: unambiguous} and Corollary \ref{cor: local-transformation}, this has consequences in our search for a universal resource: 

\begin{thm}
\label{thm: universal} Consider an $N$-partite system $\mathcal{H}=\otimes_{k=1}^{N}\mathbb{C}^{d_{k}}$ 
with $N\ge3,d_k\ge2$ for all $k$. If $\mathcal{H}$ contains more than one maximally entangled equivalence
class, then there does not exist a universal resource state $\left|\Phi\right\rangle \in\mathcal{H}$ for either local state discrimination or local unambiguous discrimination. 

In fact, there exist orthonormal bases ${\cal {B}}$ of $\mathcal{H}$
for which the set of states $\left|\Phi\right\rangle \otimes{\cal B}$
is not locally distinguishable for \textit{any} $\left|\Phi\right\rangle \in\mathcal{H}$.
\end{thm}

\begin{cor}\label{Cor: Equal Sizes}
For any $\mathcal{H}=\left(\mathbb{C}^{d}\right)^{\ot N}$ with $N\ge3,d\ge2$, there does not exist a universal resource state $\left|\Phi\right\rangle \in\mathcal{H}$ for either local state discrimination or local unambiguous discrimination. 
\end{cor}

The assumption in Theorem \ref{thm: universal} that
$\mathcal{H}$  contains a pair of incompatible maximally entangled
states is typical for multipartite spaces. For instance, in the case of three qubits, the $GHZ$ states and the $W$ states are both maximally entangled but cannot be transformed into each other, so there is no three-qubit
pure state that can optimally distinguish \emph{all} three-qubit orthonormal
bases. Any universal resource state must exist in higher
dimensions.

The second part of the theorem says that it may not be possible to find even a \emph{basis-dependent}
resource state from the same state space, i.e., there exists sets
of multipartite states for which \textit{no} multipartite pure state
from the same state space can perfectly distinguish them by LOCC.
Again in the case of three qubits, an example would be any basis which
contains at least one $GHZ$ state and one $W$ state (see \cite{DVC} and equation (\ref{eqn:WGHZ})). For such
a basis there is no three-qubit resource state which will enable perfect
state discrimination. Clearly, in this case any resource state that
perfectly distinguishes the basis using LOCC requires higher dimensions
as well.

To see the connection between Theorems 
\ref{thm: unambiguous} and \ref{thm: universal},  we assume that $\mathcal{H}$ is
a multipartite space with two distinct maximal SLOCC classes $\mathcal{C}_{1}$
and $\mathcal{C}_{2}$ with pure states $\left\vert \psi_{1}\right\rangle \in\mathcal{C}_{1}$
and $\left\vert \psi_{2}\right\rangle \in\mathcal{C}_{2}$. Since
these classes are maximal, we know that if $\left|\Phi\right\rangle \in\mathcal{H}$
and $\left|\Phi\right\rangle \rightarrow\left|\psi_{i}\right\rangle $,
then $\left|\Phi\right\rangle \in\mathcal{C}_{i}$ for $i=1,2$. Since
equivalence classes must be disjoint, there does not exist a state
$\left|\Phi\right\rangle \in\mathcal{H}$ which can be transformed
into both $\left|\psi_{1}\right\rangle $ and $\left|\psi_{2}\right\rangle $.
By Corollary \ref{cor: local-transformation}, if $\mathcal{B}_{1}$
and $\mathcal{B}_{2}$ are orthonormal bases for $\mathcal{H}$ with
$\left\vert \psi_{i}\right\rangle \in\mathcal{B}_{i}$, then it is
not possible that the sets $\left|\Phi\right\rangle \otimes\mathcal{B}_{1}$
and $\left|\Phi\right\rangle \otimes\mathcal{B}_{2}$ are both perfectly
(or even unambiguously) distinguishable with LOCC. Since this is true
for \textit{any} $\left|\Phi\right\rangle \in\mathcal{H}$, then there
is no universal resource state in $\mathcal{H}$.

Note also if we look at a basis ${\cal B}$ which contains both $\left|\psi_{1}\right\rangle $
and $\left|\psi_{2}\right\rangle $ and apply Theorem \ref{thm: unambiguous},
then the set $\left|\Phi\right\rangle \otimes{\cal B}$ cannot be
unambiguously distinguished for any $\left|\Phi\right\rangle \in\mathcal{H}$,
even if we allow $\left|\Phi\right\rangle $ to depend on ${\cal B}$.

As was previously observed, if we allow $\left|\Phi\right\rangle $
to live in a higher-dimensional space, then it might be possible to
convert into different maximal SLOCC classes of $\mathcal{H}$. While characterizing universal resource states for perfect discrimination
has proved challenging, our results give necessary and sufficient
conditions for unambiguous discrimination as a corollary to Theorem
\ref{thm: unambiguous}: 
\begin{cor}\label{Cor: Unambiguous Universal}
Let $\mathcal{H}=\otimes_{k=1}^{N}\mathbb{C}^{d_{k}}$ be a multipartite
system and let $\mathcal{H'}=\otimes_{k=1}^{N}\mathbb{C}^{d'_{k}}$
be a system with the same multipartite structure as $\mathcal{H}$.

Then $\vert\Phi\ket\in\mathcal{H'}$ is a universal resource for 
unambiguous discrimination if and only if for every maximally entangled
state $\vert\phi\ket\in\mathcal{H}$, $\vert\Phi^*\ket$ can be locally
transformed into $\vert\phi\ket$. \end{cor} 
That is, to test whether a state is universal for unambiguous discrimination in $\mathcal{H}$, one need only test whether it can be transformed into each of the maximally-entangled states of $\mathcal H$. (Characterizing the set of maximally-entangled states was the focus of recent work in \cite{HSK-2015}.)
The 
corollary follows since, for every $\vert\phi'\ket\in{\mathcal{H}}$,
there exists a maximally entangled state $\vert\phi\ket$ which can
be locally transformed into $\vert\phi'\ket$. Hence, $\vert\Phi^*\ket\rightarrow\vert\phi\ket\rightarrow \vert \phi'\ket$,
and we can apply Theorem \ref{thm: unambiguous}.

\section{Proof of Theorem \ref{thm: unambiguous}} \label{Section:Proofs}

We begin by showing that local unambiguous discrimination implies local transformation with fixed probabilities of success. The following lemma is easily checked: 
\begin{lem}\label{IdentityLemma}
Let $\{ \vert \psi_i \ket \}$ be a complete basis of $\mathcal{H}$, not necessarily orthogonal; and for each $i$, let $\vert \tilde{\psi}_i \rangle$ be the unique unit vector in $\mathcal{H}$ such that $\langle  \tilde{\psi}_i \vert \psi_j \rangle = 0$ if $i \ne j$. 

If $\dim D = \mathcal{H} $, $\mathcal I$ is the identity operator on $\mathcal{H}$ and $\vert \Psi \rangle$ is the standard maximally-entangled state on $\mathcal{H} \ot \mathcal{H}$, then we can write
\bee
 \mathcal{I} = \sum_i \frac{1}{\langle  \tilde{\psi}_i \vert \psi_i \rangle } \vert \psi_i \rangle\langle \tilde{\psi}_i \vert 
\eee and
\bee
 \vert \Psi \ket = \frac{1}{\sqrt{D}} \sum_i \frac{1}{\langle  \tilde{\psi}_i \vert \psi_i \rangle } \vert \psi_i^* \rangle\vert \tilde{\psi}_i \rangle  
\eee
where $\vert \psi_i^* \rangle$ is the entrywise complex conjugate of $\vert \psi_i \rangle$. 
\end{lem}
Now suppose that there exists an
LOCC measurement $\boldsymbol{\Pi}=\left\{ \Pi_{i}\right\} _{i=1}^{D}$
satisfying ${\displaystyle \sum_{i}\Pi_{i}={\cal I}_{\mathcal{H}^\prime \ot \mathcal{H}}}$
such that 
\begin{eqnarray}
\left\langle \Phi\otimes\psi_{j}\left|\Pi_{i}\right|\Phi\otimes\psi_{j}\right\rangle  & = & \delta_{i,j}\epsilon_i \end{eqnarray}
for $\epsilon_i > 0$.  We show that there exists a local protocol which effects the transformation $\vert \Phi^* \ket \rightarrow \vert \psi_i \ket$ with probability $\frac{\epsilon_i}{D}$ for each $i$. 

Suppose that  $N$ spatially-separated observers $\{O_{k}\}$ each control a subsystem of $\mathcal{H}^\prime$ and initially share the state $\left\vert \Phi^*\right\rangle \in \mathcal{H}^\prime$. Each observer $O_k$ locally produces the maximally entangled state $\vert \Psi_k \ket \in \mathbb{C}^{d_k} \ot \mathbb{C}^{d_k}$ in an ancillary system. We can then write the maximally entangled state on $\mathcal{H} \ot \mathcal{H}$ as $\vert \Psi \ket = \otimes_{k=1}^{N} \vert\Psi_{k}\ket$. This means that the state of our entire system is given by
\bee
\vert \Phi^* \ket \ot  \vert \Psi \ket \in \mathcal{H}^\prime \ot \left( \mathcal{H} \ot  \mathcal{H}\right)
\eee
Using lemma \ref{IdentityLemma}, we can write
\bee
\vert \Psi \ket & = & \frac{1}{\sqrt{D}} \sum_i \frac{1}{\langle  \tilde{\psi}_i \vert \psi_i \rangle } \ot\vert \psi_i^* \rangle\vert \tilde{\psi}_i \rangle \\
\vert \Phi^* \ket \ot \vert \Psi \ket & = & \frac{1}{\sqrt{D}} \sum_i  \frac{1}{\langle  \tilde{\psi}_i \vert \psi_i \rangle }\vert \Phi^* \ket  \ot  \vert \psi_i^* \rangle\ot\vert \tilde{\psi}_i \rangle \eee

Notice that we have created our copy of $\vert \Psi \ket$ using only local operations on our $N$ subsystems. Now, we take the entrywise conjugate of each operator of our LOCC measurement $\boldsymbol{\Pi}$ and apply the new measurement $\boldsymbol{\Pi}^*$ to the first two systems. If we get the outcome $x$, then our system has been transformed as
\bee
(\sqrt{\Pi_x^*} \ot \mathcal{I})\vert \Phi^* \ket  \vert \Psi \ket & = &  \frac{1}{\sqrt{D}} \sum_i \sqrt{\Pi^*_x}\left(\vert \Phi^* \ket \vert \psi_i^* \ket \right) \ot \vert \psi_i \ket \\
& =&  \frac{1}{\sqrt{D}} \sqrt{\Pi^*_x}\left(\vert \Phi^* \ket \vert \psi_x^* \ket \right) \ot \vert \psi_x \ket 
\eee
Tracing out the first two systems shows that our last system ends up in the state $\vert \psi_x \ket$ with probability $\frac{1}{D} \left\langle \Phi\otimes\psi_{x}\left|\Pi_{x}\right|\Phi\otimes\psi_{x}\right\rangle = \frac{\epsilon_{x}}{D} >0$, which was to be shown.

 To prove the  converse, we assume that
the LOCC transformation $\vert\Phi^{*}\ket\rightarrow\vert\tilde{\psi_{i}}\ket$
is possible for each $i$. This implies that there exists a product
matrix $M_{i}={\displaystyle \ot_{k=1}^{N}M_{i}^{(k)}}$ with $\tr M_i^*M_i = D'$ and $\vert\tilde{\psi_{i}}\ket=\mu_i M_{i}\vert\Phi^{*}\ket$ for some  constant $\mu_i$. We will use this to build a measurement which unambiguously distinguishes
the $\{\vert\psi_{i}\ket\}$.

For each subsystem $k$, let $\vert\Psi_{k}^\prime\ket$ be the standard maximally-entangled   $d_k^\prime \ot d_k^\prime$ state so that $\vert \Psi^\prime \ket = \otimes_{k=1}^{N} \vert\Psi_{k}^\prime\ket$ is maximally entangled on $\mathcal{H}^\prime \ot \mathcal{H}^\prime$. For each $i$, we can also define $\vert\phi_{i}\ket=(I\ot M_{i})\vert\Psi^\prime\ket$, which is a product state across our $N$ subsystems. This allows us to write
\begin{eqnarray*}
\vert\phi_{i}\ket & = & \ot_{k=1}^{N} \left(I \ot M_{i}^{(k)} \vert \Psi_k^\prime\ket\right)  \\&=& (I\ot M_{i})\vert\Psi^\prime\ket \\
\left\langle \left(\Phi\otimes\psi_{j}\right)\vert\phi_{i}\right\rangle  & = & \left\langle \left(\Phi\otimes\psi_{j}\right)\left|\left(I\otimes M_{i}\right)\right|\Psi^\prime\right\rangle\\
 & = & \frac{1}{\sqrt{D'}}\left\langle \psi_{j}\left|M_{i}\right|\Phi^{*}\right\rangle =\frac{1}{\sqrt{D'}}\left\langle \psi_{j}\vert\tilde{\psi_{i}}\right\rangle 
\end{eqnarray*}
By definition, $\left\langle \left(\Phi\otimes\psi_{j}\right)\vert\phi_{i}\right\rangle  = \left\langle \psi_{j}\vert\tilde{\psi_{i}}\right\rangle \ne0$
if and only if $i=j$. Since the  $\{ \vert \phi_i \ket \}$ are product states, Lemma 3 \cite{Duan2007} establishes that they can be used to  unambiguously distinguish the states $\{ \vert \Phi \ot \psi_i \ket \}$ using only LOCC. 
\qed

\section{Examples}\label{section:Examples}
When we look for universal resources in multipartite systems, we can mimic the bipartite structure to identify an example of a universal resource state: If one of the subsystems shares sufficient bipartite
entanglement with each of the other subsystems, then teleportation can be used to recreate the entire unknown state in one location, after
which discrimination is possible. 
\begin{exm} \label{exm: Bell} Let $\mathcal{H}=\otimes_{k=1}^{N}\mathbb{C}^{d_{k}}$
be a multipartite system with $D = \dim {\cal H} = d_1d_2\cdots d_N$. We define a state in which each of the first $(N-1)$ subsystems shares a maximally-entangled state with the $N$th system: 
\bee
\left|\Phi_{Bell}^{N}\right\rangle := \sqrt{\frac{d_N}{D}} \sum_{i= 1}^{D/d_N} \vert i \ot i \ket \in \left(\otimes_{k=1}^{N-1}\mathbb{C}^{d_{k}}\right)\otimes\mathbb{C}^{D/d_{N}}
\eee
Then $\left|\Phi_{Bell}^{N}\right\rangle$ is a universal
resource for both local unambigious discrimination and local perfect discrimination in $\mathcal{H}$. 
\end{exm}
Note that in general this requires
the resource state to exist in a much higher-dimensional system that
our original states, since the dimension of the last subsystem is the product of the dimensions of all the other subsystems. 

This is not the only possible resource state, however.  Corollary \ref{Cor: Unambiguous Universal} implies that $\left|\Phi\right\rangle $ is a universal resource
for unambiguous discrimination if and only if its SLOCC class is an
upper bound for every SLOCC class in $\mathcal{H}$. This requires $\vert \Phi \ket$ to possess sufficient entanglement, as measured by any entanglement monotone. In particular, we can look at the Schmidt measure
\be
E_S(\vert \Phi \ket) := \log r
\ee
where $r$ is the minimum number of terms in any representation of $\vert \Phi \ket$ as a linear combination of product states. In bipartite systems, $r$ is simply the rank of the reduced density matrix, but $E_S$ is well-defined as an entanglement monotone in multipartite systems as well \cite{DVC,EB2001,HEB, HM}. The Schmidt rank of a pure state cannot increase under LOCC , which
implies that for any universal resource state $\left|\Phi\right\rangle $,
$E_S\left(\vert\Phi\ket\right)\ge E_S\left(\vert \psi \ket \right)$ for all $\left|\psi\right\rangle \in\mathcal{H}$.
This necessary condition leads us to a second class of universal resource
states for unambiguous discrimination:

\begin{exm} \label{exm: GHZ} Let $\mathcal{H}=\otimes_{k=1}^{N}\mathbb{C}^{d_{k}}$
be a multipartite system. Following the notation in \cite{CCDJW-2010}, we define $\left|GHZ_N^R\right\rangle \in\left(\mathbb{C}^{R}\right)^{\otimes N}$
to be the generalized GHZ state in dimension $R$ with $N$ parties:
\begin{eqnarray*}
\left|GHZ_N^R\right\rangle  & = & \frac{1}{\sqrt{R}}\sum_{i=0}^{\left(R-1\right)}\left|i\right\rangle ^{\otimes N}
\end{eqnarray*}

Then $\left|GHZ_N^R\right\rangle $ is a universal
resource for unambiguous discrimination in $\mathcal{H}$ if and only if 
\bee
\log R \ge  \max_{\vert \psi\ket \in \mathcal{H}} E_S(\vert \psi \ket)
\eee \end{exm}

The necessity follows from the monotonicity of the Schmidt measure. The sufficiency is given as Observation 1 in \cite{CCDJW-2010}, which establishes that this state can be transformed into any state with smaller Schmidt rank; and the specific case when $\mathcal{H} = \left(\mathbb{C}^2\right)^{\ot 3}$ is proved in \cite{GCDResource2016}. The proof is immediate: We can write any state 
$\left|\psi\right\rangle \in\mathcal{H}$ as a linear combination of at most $R$ product states: 
\begin{eqnarray*}
\left|\psi\right\rangle  & = & \sum_{i=1}^{R}\alpha_{i}\otimes_{k=1}^{N}\left|a_{i,k}\right\rangle \end{eqnarray*}
If we define  $A_{k}=\sum_{i}\left(\alpha_{i}\right)^{1/N} \vert a_{i,k}\rangle\langle i\vert$, then we can write 
\bee
\left|\psi\right\rangle  & = & \sqrt{R} \left(\otimes_{k=1}^{N}A_{k}\right)\left|GHZ_N^R\right\rangle
\eee
which shows that we can locally transform $\left|GHZ_N^R\right\rangle$ into any state $\left|\psi\right\rangle \in \mathcal{H}$.

The  maximum rank $R$ of $\mathcal{H}$ defined in this  example is not easily calculated. However, the example of generalized $W$ states in \cite{TensorRank2009, CCDJW-2010} give us that if $d = 2^n$ and $\mathcal{H} = \left(\mathbb{C}^d\right)^{\ot N}$, then $R \ge (N-1)(d-1)+1$. 

This now gives us two classes of universal resources which are opposite
extremes: $\left|\Phi_{{\rm Bell}}^{N}\right\rangle $ has minimal
dimension in all parties except one, while $\left|GHZ_N^R\right\rangle $
has uniform dimensions across each party and the minimum dimension
for which this is possible. For instance, in the case of three qubits,
the minimum rank is $R=3$, so $\left|GHZ_N^R\right\rangle $
lives in a $3\otimes3\otimes3$ system, while $\left|\Phi_{{\rm Bell}}^{N}\right\rangle $
lives in a $4\otimes2\otimes2$ system. It is an open question to
characterize \textit{all} universal resource states for a fixed system
$\mathcal{H}$.

Note that $\left|\Phi_{{\rm Bell}}^{N}\right\rangle $ is clearly
a universal state for perfect discrimination as well as unambiguous
discrimination, while it is not clear whether this is true for $\left|GHZ_N^R\right\rangle $.
The following example shows that one need not imply the other: \begin{exm}
\label{ex: 3 qubit} Consider the three-qubit system $\mathcal{H}=\mathbb{C}^{2}\otimes\mathbb{C}^{2}\otimes\mathbb{C}^{2}$
and the resource state 
\begin{eqnarray}
 & \left|\Phi\right\rangle =\frac{1}{\sqrt{3}}\left(\left|000\right\rangle +\left|110\right\rangle +\left|201\right\rangle \right)\in\mathbb{C}^{3}\otimes\mathbb{C}^{2}\otimes\mathbb{C}^{2}
\end{eqnarray}
Then $\left|\Phi\right\rangle $ is universal for the problem of unambiguous
discrimination in $\mathcal{H}$ but not perfect state discrimination.
\end{exm}

In order to be universal for unambiguous state discrimination in $\mathcal{H}$,
we need only show that $\left|\Phi\right\rangle $ can be transformed
into both a $W$ state and a $GHZ$ state. We can transform one state
into another if they are related by a product matrix, and it is not
hard to see that 
\begin{eqnarray}
\left|W\right\rangle  & = & \frac{1}{\sqrt{3}}\left(\left|100\right\rangle +\left|010\right\rangle +\left|001\right\rangle \right) \label{eqn:WGHZ} \\
 & = & \left(\left[\begin{array}{ccc}
0 & 1 & 1\\
1 & 0 & 0
\end{array}\right]\otimes\mathcal{I}_{2}\otimes\mathcal{I}_{2}\right)\left|\Phi\right\rangle \nonumber
\end{eqnarray}
\begin{eqnarray*}
\left|GHZ\right\rangle  & = & \frac{1}{\sqrt{2}}\left(\left|000\right\rangle +\left|111\right\rangle \right)\\
 & = & \sqrt{\frac{3}{2}}\left(\left[\begin{array}{ccc}
0 & 1 & 1\\
1 & 0 & 0
\end{array}\right]\otimes\sigma_{{\rm x}}\otimes\mathcal{I}_{2}\right)\left|\Phi\right\rangle 
\end{eqnarray*}
This means that $\left|\Phi\right\rangle $ can be transformed into
$\left|W\right\rangle $ and $\left|GHZ\right\rangle $. These are
the two maximally entangled three-qubit SLOCC classes, and Corollary
\ref{Cor: Unambiguous Universal} tells us that this is sufficient
for $\vert\Phi\ket$ to be universal for unambiguous discrimination.

On the other hand, if $\mathcal{B}$ is a basis of $\mathcal{H}$
which consists entirely of $W$ states and $GHZ$ states (including at least
one $GHZ$ state), then the set $\left|\Phi\right\rangle \otimes\mathcal{B}$
cannot be perfectly distinguished with LOCC. This can be seen by looking
at the bipartite entanglement across the $AB:C$ split. Split this
way, $\left|W\right\rangle $ and $\left|\Phi\right\rangle $ are
in the same bipartite SLOCC class of $\mathbb{C}^{6}\otimes\mathbb{C}^{2}$,
and they possess $H\left(\frac{2}{3}\right)<1$ units of bipartite
entanglement, while the $GHZ$ state has a full unit $H\left(\frac{1}{2}\right)=1$
of $AB:C$ entanglement.

Since average bipartite entanglement cannot increase under LOCC, there
does not exist an LOCC protocol which transforms $\left|\Phi\right\rangle $
into each element of $\mathcal{B}$ with probability $\frac{1}{8}$; any
protocol which sometimes gains entanglement by transforming $\left|\Phi\right\rangle $
into $\left|GHZ\right\rangle $ must also sometimes lose entanglement
as well. This result implies that the elements of $\left|\Phi\right\rangle \otimes\mathcal{B}$
are not perfectly distinguishable with LOCC, according to Corollary \ref{cor: local-transformation}.

\section{Conclusion}\label{section:Conclusions}

We have shown that for a fixed multipartite system there
is often no resource from the same state space that can enable
all complete orthogonal measurements on the whole system by LOCC. This is always the case when the dimensions of the $N \ge 3$ subsystems
are all equal. This is in sharp
contrast to the bipartite scenario, where a maximally entangled state
of full Schmidt rank serves as a universal resource state. Furthermore,
there exist orthonormal bases for which one cannot find any resource
state from the same state space that perfectly distinguishes the basis
states. This property of multipartite spaces is found to be typical
even if we allow the dimensions of the subsystems to be different
from one another; exceptions arise in scenarios where dimension of
one of the subsystems is much larger than the dimensions of the other
subsystems.

This line of questioning suggests many open problems. There is much that is not known about multipartite entanglement and the structure of SLOCC entanglement classes; and trying to understand the nature of universal resource states gives a possible line of approach. Certainly, it
would be useful to find a complete characterization of universal resource
states, perhaps finding ways to adapt the methods in \cite{GW2011}
to do so. Even just bounding the necessary dimensions for such states would be a step in the right direction. The strength of Theorem \ref{thm: unambiguous} makes it seem like the search for universal states for unambiguous local discrimination would be the most promising. One could also try to determine whether the
generalized GHZ states in Example \ref{exm: GHZ} are universal
for perfect state discrimination. This is related to the question of finding optimal resources in \cite{GCDResource2016}; or even knowing whether a unique optimal resource exists. A larger question
would be to find efficient universal resources to accomplish \textit{any}
quantum operation as efficiently locally as one could globally. For
this, a {\it pair} of bell states $\vert\Phi_{Bell}^{N}\ket$ could suffice
(one to teleport everything into one place, and one to teleport them
back); but it would be nice to have a smaller resource for this. It
is hoped that this line of questioning will enable the continued exploration
of the interplay between locality and entanglement.
\begin{acknowledgments}
SB is supported in part by DST-SERB project SR/S2/LOP-18/2012. SH
is supported by a fellowship from CSIR, Govt. of India. MN acknowledges the continued support of the Office of Faculty Development at Saint Mary's College. \end{acknowledgments}


\end{document}